\documentclass[12pt]{article}
\pdfoutput=1
\usepackage{epsfig}
\usepackage{amsfonts}
\usepackage{amssymb}
\usepackage{xcolor}
\usepackage{epsfig}
\usepackage{amsfonts}
\usepackage{amssymb}
\usepackage{xcolor}
\usepackage{mathrsfs}
\usepackage{amsmath}
\usepackage{upgreek}
\usepackage{graphicx}
\usepackage{cite}
\usepackage{calligra}
\usepackage{caption}
\usepackage{subcaption}
\usepackage{cite}
\usepackage{calligra}
\usepackage[english]{babel} 
\usepackage{hyperref}

\usepackage{bm}
\usepackage{amssymb}
\usepackage{lipsum}
\begin{document}
\newcommand{\nc}{\newcommand}
\newcommand{\kms}{km\,s$^{-1}$}
\newcommand{\msun}{$M_\odot}
\nc{\beq}{\begin{equation}} \nc{\eeq}{\end{equation}}
\nc{\beqa}{\begin{eqnarray}} \nc{\eeqa}{\end{eqnarray}}
\nc{\R}{{\cal R}}
\nc{\A}{{\cal A}}
\nc{\K}{{\cal K}}
\nc{\B}{{\cal B}}
\nc{\ba}{\begin{array}} \nc{\ea}{\end{array}}
\renewcommand*\thefootnote{\alph{footnote}}
\def\gm{\mathfrak{g}}
\def\Up{\Upsilon}
\def\hc{\ensuremath{\mathrm{h.c.}}}
\begin{center}
{\bf \Large  Leading UV divergences of quantum corrections to \\ Kähler superpotential in general $\mathcal{N}=1$ chiral model} \vspace{1.0cm}

{\bf \large R.M. Iakhibbaev$^{1,a}$, A. I. Mukhaeva$^{1,b}$ and D.M. Tolkachev$^{1,2,c}$} \vspace{0.5cm}

{\it $^1$Bogoliubov Laboratory of Theoretical Physics, Joint Institute for Nuclear Research,
  6, Joliot Curie, 141980 Dubna, Russia \\and \\
$^2$Stepanov Institute of Physics,
68, Nezavisimosti Ave., 220072, Minsk, Belarus}
\vspace{0.5cm}

\abstract{Using the Bogoliubov-Parasiuk theorem we derive differential equations for the sum of leading UV divergences of the Kähler potential in the general  $\mathcal{N}=1$  supersymmetric chiral theory. The obtained equations recover the limit of the renormalizable Wess-Zumino theory and also allow one to consider non-renormalizable chiral interactions. Some implications of the obtained equations are shown.}
\end{center}

\text{\footnotesize{ E-mails: $^a$yaxibbaev@jinr.ru, $^b$mukhaeva@theor.jinr.ru, $^c$dtolkachev@jinr.ru,
}}

\text{\footnotesize{Keywords: effective potential, supersymmetry, non-renormalisable theories, renormalisation group}}

 \section{Introduction}
It is an open question as to which approaches are more appropriate for studying of non-renormalizable models. Nevertheless, recent progress in studying  non-renormalizable theories associated with the BPHZ procedure \cite{BP,Hepp,Zimmermann} made it possible to obtain recurrence relations and (integro)differential equations connecting the divergent parts of the Feynman loop integrals in different orders \cite{we2015,we2019}. These equations are generalizations of the usual renormalization-group equations and were applied to study effective potentials in scalar theories and amplitudes in various models \cite{we2015,we2022,we2019,Kazakov:2022pkc}. Although the issue of the ambiguity of the subtraction procedure is not fully clarified, computation and summation contributions of different orders predicting the structure of  divergences is currently possible \cite{we2019,Kazakov:2020kbj,Kazakov:2022pkc,Iakhibbaev:2026fst}. In this work, we consider leading divergences of  the effective Kähler potentials, since, as is known, leading divergences (as well as leading logarithms) are known to be independent of the subtraction scheme. 

The model under consideration has the following Lagrangian:
\beq
\mathcal{L}=\int d^2 \theta d^2 \bar{\theta} ~ K(\Bar{\Phi},\Phi)+ \int d^2 \theta ~ \lambda W(\Phi)+ \hc, 
\label{GenLagr}
\eeq
where $\theta^\alpha, \bar{\theta}_{\dot{\alpha}}$ are Grassman variables, $K$ is an arbitrary function of superfields, $\lambda W(\Phi)$ is an arbitrary superfield interaction potential with $\lambda$ being a real coupling constant, $\Phi(z)$ and $\bar{\Phi}(z)$ are chiral and antichiral superfields depending on $\mathcal{N}=1$ superspace coordinates $z^M=(x^\mu, \theta^\alpha,\bar{\theta}_{\dot{\alpha}})$ with $x^\mu$ being usual Minkowski space coordinates. The general chiral-superfield model considered here is widely used to study possible phenomenological implications of superstring theory \cite{Green:1987sp,Buchbinder:1999sk,Buchbinder:1999jw}. Notice that $K=\Phi \bar{\Phi}$ and $W(\Phi)=\Phi^3/3!$ correspond to the usual Wess-Zumino model, $W(\Phi)=0$ corresponds to the supersymmetric sigma model \cite{Buchbinder:1998twe}. Recall, the mass dimensions of (anti)chiral superfields $[\Phi]=[\bar{\Phi}]=1$, and for the Grassman variables the mass dimension reads $[d \bar{\theta}]=[d \theta]=[ \theta]=[ \bar\theta]=1/2$.

It is known that the non-renormalization theorem strongly restricts the number of divergent structures \cite{Buchbinder:1993ud,Buchbinder:1998twe}. In the model under consideration only two-point structures are UV divergent. Moreover, in the effective potential the divergences are generated only by corrections to the Kähler potential\footnote{It should be noted that the effective potential in supersymmetric theories can consist of three parts: the Kähler effective potential (containing UV divergences), as well as the effective potential of auxiliary fields and the chiral potential (containing only finite contributions) \textcolor{black}{The auxiliary field effective superpotential is the component of the effective superspace action that depends on spinor derivatives of the superfields and vanishes when those derivatives are set to zero \cite{Buchbinder:1994xq,Kuzenko:2014ypa}. The chiral effective potential is the finite part of the effective action that occurs exclusively in massless renormalizable theories (see details in Refs.\cite{Buchbinder:1993ud,Buchbinder:1994xq,Buchbinder:1998twe,Kuzenko:2014ypa}).}}, since singular perturbative corrections to chiral potentials $W(\Phi)$ are forbidden because a quantum coorrection should be an integral over the full superspace \cite{Buchbinder:1993ud,Buchbinder:1994df,Buchbinder:1999ui}. To proceed, we obtain the Feynman rules and calculate first two-point UV-divergent graphs contributing to the Kähler effective potential.

\begin{figure}
    \centering
    \includegraphics[width=1.0\linewidth]{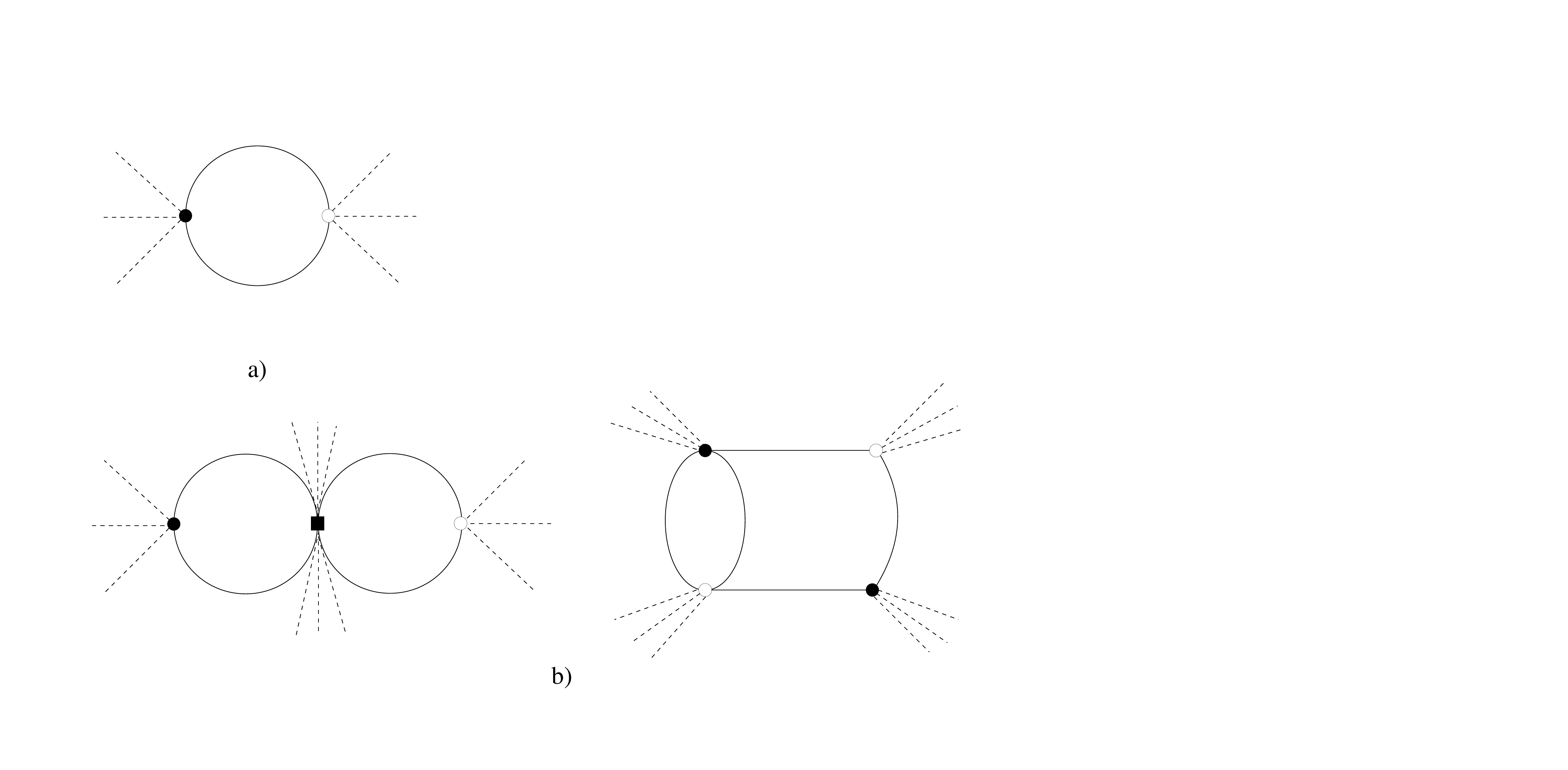}
    \caption{Contributions to the Kähler effective potentials: a) one-loop and b) two-loop Feynman diagrams. The dashed lines denote zero-momentum classical lines of $\Phi$ and $\bar{\Phi}$. The white and black blobs are the vertices corresponding to covariant derivatives over the chiral superpotential and the square dot is a vertex containing derivatives over the Kälher potential.}
    \label{fig:graphs}
\end{figure}

A more direct way to perturbatively calculate the effective potential is to sum all 1PI vacuum superdiagrams (diagrams with zero external momentum \footnote{The effective potential should be evaluated at the superfield conditions
$\partial_\mu \Phi = \partial_\mu \bar{\Phi}=0$ \cite{Buchbinder:1998twe}, i.e. we should extract part of the effective action that does not depend on momenta}) obtained using the Feynman rules derived from the shifted action, as shown in Refs. \cite{Abbott:1981ke,Buchbinder:1999ui}. Let $\Gamma [\bar{\Phi},\Phi]$ be an effective action in the model \eqref{GenLagr}.
Due to the shift $\Phi \rightarrow \Phi+\sqrt{\hbar}\phi$, where $\Phi$ represents the classical field, the effective action is given by
\begin{eqnarray}
\label{Green1}
 e^{\frac{i}{\hbar}\Gamma[\bar{\Phi},\Phi]} &=&
 \int {\cal D} \phi {\cal D} \bar{\phi}
~\exp\big(
\frac{i}{\hbar}
S[\bar{\Phi}+\sqrt{\hbar}\bar{\phi},\Phi+\sqrt{\hbar}\phi]
-\nonumber\\&-&
\left(\int d^6 z
\frac{\delta\Gamma[\bar{\Phi},\Phi]}{\delta\Phi(z)}\phi(z)+h.c. \right)
\big).
\end{eqnarray}
To find propagators and vertices from $\Gamma[\bar{\Phi},\Phi]$ loop expansion in explicit form, we expand the right-hand side of (\ref{Green1}) in series in quantum superfields $\phi$, $\bar{\phi}$. As usual, the quadratic part of the expansion of
$S[\bar{\Phi}+\sqrt{\hbar}\bar{\phi},\Phi+\sqrt{\hbar}\phi] $
\begin{eqnarray}
\label{S2action}
S_2=\frac{1}{2}\int d^8 z \left(\begin{array}{cc}\phi&\bar{\phi}
\end{array}\right)
\left(\begin{array}{cc}
K_{\Phi\Phi}&K_{\Phi\bar{\Phi}}\\
K_{\Phi\bar{\Phi}}&K_{\bar{\Phi}\bar{\Phi}}
\end{array}\right)
\left(\begin{array}{c}\phi\\
\bar{\phi}
\end{array}
\right)+\\+\left(\int d^6 z \frac{\lambda}{2}W^{''}\phi^2+h.c.\right)
\end{eqnarray}
defines the propagators. Higher terms of the effective action expansion define the vertices \cite{Buchbinder:1994iw,Buchbinder:1994xq,Buchbinder:1999ui,Buchbinder:1999sk}. Here $K_{\bar{\Phi}\Phi}$ is the corresponding derivative of the classical $K(\bar{\Phi},\Phi)$ and defines the Kählerian metric $\gm$.
A matrix superpropagator between two points $z_1,z_2$ in the superspace can be defined using a quadratic action:
\beq
\label{matrixprop}
 G(z_1,z_2) =
 \left(
 \begin{array}{ll}
  G_{++}(z_1,z_2) & G_{+-}(z_1,z_2)\\
  G_{-+}(z_1,z_2) & G_{--}(z_1,z_2)
 \end{array}
 \right)
\eeq
so one can obtain equation of motion for the propagator from the quadratic action:
\beq
\label{def}
 \left(
 \begin{array}{cc}
  W''-\frac{1}{4}(\bar{D}^2 K_{\Phi\Phi}) &-\frac{1}{4}\bar{D}^2 K_{\Phi\bar{\Phi}}\\
-\frac{1}{4}D^2  K_{\Phi\bar{\Phi}} &
\bar{W''}-\frac{1}{4}(D^2 K_{\bar{\Phi}\bar{\Phi}})
 \end{array}
 \right)
 \left(
 \begin{array}{ll}
  G_{++} & G_{+-}\\
  G_{-+} & G_{--}
 \end{array}
 \right)
 =
-\left(
 \begin{array}{ll}
  \delta_+ & 0\\
  0 & \delta_-
 \end{array}
 \right)
\eeq
where $\delta_+=-\frac{1}{4}\bar{D}^2\delta^8(z_1-z_2)$,
$\delta_-=-\frac{1}{4}D^2\delta^8(z_1-z_2)$ are chiral and antichiral delta functions, respectively, and $D$-operators of non-diagonal elements act on the all function at the
rhs.
Thus, the components of the matrix superpropagator in the case of constant superfields are given as 
\beqa
G_{++}&=&\frac{\lambda \bar{W}''}{K_{\Phi\bar{\Phi}}\square - \lambda^2 |W''|^2} \frac{D_1^2}{16}\delta_{12}\label{Dprop1},\\ ~ G_{--}&=&\frac{\lambda W''}{K_{\Phi\bar{\Phi}}\square - \lambda^2 |W''|^2} \frac{\bar D_1^2}{16}\delta_{12},
\label{Dprop2}
\eeqa
and
\beqa
G_{+-}&=&\frac{1}{K_{\Phi\bar{\Phi}}\square - \lambda^2 |W''|^2} \frac{D_1 \bar{D}_2}{16}\delta_{12} \label{NDprop1}, \\~ G_{-+}&=&\frac{1}{K_{\Phi\bar{\Phi}}\square - \lambda^2 |W''|^2} \frac{\bar{D}_1 D_2}{16}\delta_{12} \label{NDprop2}.
\eeqa
Note that the effective mass $m^2=\lambda^2 |W''|^2$ emerges due to the shifted expanded action \eqref{S2action}, and propagators contain  an infinite number of classical field insertions \cite{Pickering:1996he}. Note that if one consider \textcolor{black}{Wess-Zumino like models}, the diagrams containing the propagators $G_{++}$ from formula \ref{Dprop1} can connect only vertices proportional to $(-D/4)$ and $(-D/4)$ i.e., connect two chiral vertices  (and vice versa for $G_{--}$).
Diagrams with propagators like these cannot contribute to the one-loop Kähler effective potential, i.e. in calculations one should use the propagators from equations (\ref{NDprop1}-\ref{NDprop2}) which generates sunset-type graphs. However, due to the fact that the Kähler potential can be an arbitrary function, bubble-type diagrams (i.e. '8'-shaped graphs, see Fig. \ref{fig:graphs}) can also contribute to the effective Kähler potential; moreover, the sunset-type diagrams become more complicated (see discussions in Refs. \cite{GrootNibbelink:2005nez,Buchbinder:1998twe}). 

The vertices (which can be obtained from the expansion of chiral superpotential) are generally expressed as derivatives of the chiral potential in the following form:
\beq
    w_j=\frac{\partial^j W}{\partial \Phi^j},~\bar{w}_j=\frac{\partial^j \bar{W}}{\partial \bar{\Phi}^j},
\eeq
so now we denote $|W''|^2\equiv|w_2|^2$.
Part  of the Kähler effective potential can be written as a formal expansion:
\beq
\mathcal{K}_{eff}=\sum_{j=0}^\infty (\lambda^{2})^j \mathcal{K}_j(\bar{\Phi},\Phi), \label{sumK}
\eeq
where $\mathcal{K}_0=K(\bar{\Phi},\Phi)$. To simplify the loop calculation results, it is convenient to introduce new shorthand notation for affine connections $\gamma=\gm^{-1}K_{\bar{\Phi}\Phi\Phi}$, $\bar{\gamma}=\gm^{-1}K_{\bar{\Phi}\bar{\Phi}\Phi}$ and for the useful variable $\rho=K_{\bar{\Phi}\Phi\bar{\Phi}\Phi}-\bar{\gamma} \gm \gamma$ which is a curvature tensor on the Kähler manifold.
In this notation, using dimensional regularization $d=4-2\epsilon$,  the first singular contribution can be represented as 
\beq
\mathcal{K}_1=-\frac{\lambda^2}{(4 \pi)^2}\int d^4\theta~ k_1[\gm] \left(\frac{1}{\epsilon}+O(\epsilon^0)\right), \label{oneloop}
\eeq
where $k_1=|w_2|^2/2\gm^2$, and the factor $\gm^{-2}$ originates from the propagators \ref{Dprop1}-\ref{NDprop2}. This result is in agreement with the one-loop calculations in \cite{Buchbinder:1998twe,Pickering:1996he,GrootNibbelink:2005nez}
The second loop  leading singular contribution has the following form:
\beq
\mathcal{K}_2=-\frac{\lambda^4}{(4 \pi)^4}\int d^4\theta ~k_2[\gm] \left(\frac{1}{\epsilon^2}+O(\epsilon^0)\right), \label{twoloop}
\eeq
where the coefficient is given by the sum of different geometric contributions $k_2=c_1+c_2+c_3$ 
\begin{equation}
\begin{gathered}
    c_1=\frac{|w_2|^2 |w_3|^2}{4\gm^5}, ~~c_2=-\frac{w_2^2 \bar{w}_2^2}{2 \mathfrak{g} ^6}  (\rho-2 \gamma \mathfrak{g} \bar{\gamma } ),\\
    c_3=-\frac{|w_2|^2}{2 \mathfrak{g} ^5} \left(\gamma  w_2 \bar{w}_3+\bar{\gamma } w_3 \bar{w}_2\right).
\end{gathered}
\end{equation}
The first contribution is the result of purely 'sunset' diagram calculation, and $c_2,c_3$ arises from both sunset and '8'-type diagrams \ref{fig:graphs}.

These expressions are covariant and consistent with the results found in Ref. \cite{GrootNibbelink:2005nez} after appropriate substitutions and restoring the corresponding symmetrical factors. In the case of the massless Wess-Zumino model we have the known results in the minimal substraction scheme for the leading contribution to the one-loop Kähler effective potential up to $O(\epsilon^0)$ terms:
\beq
\mathcal{K}_1^{WZ}=-\frac{\lambda^2}{(4 \pi)^2}\int d^4\theta~ \frac{\bar{\Phi}\Phi}{2\epsilon}, \label{oneloopWZ}
\eeq
and for the two-loop Kähler effective potential up to $O(\epsilon^{-1})$ terms
\beq
\mathcal{K}_2^{WZ}=-\frac{\lambda^4}{(4 \pi)^4}\int d^4\theta~ \frac{\bar{\Phi}\Phi}{4\epsilon^2}.  \label{oneloopWZ}
\eeq
These results are in agreement with numerous calculations performed in, e.g., \cite{Huq:1977eu,Grisaru:1982df,Buchbinder:1993ud,Buchbinder:1994df,Buchbinder:1994iw,Pickering:1996he,Buchbinder:1999sk,Martin:2001vx,Martin:2024qmi}. Let us note once again that the leading poles and leading logarithms have the same coefficients and do not depend on the counteterm subtraction procedure \cite{Kazakov:2022pkc}.

In this renormalizable case, the leading corrections obey the Ovsyannikov-Callan-Symanzik equation, so the renormalization group equation can be obtained.  The renormalization group equations of the Wess-Zumino model are known and can be found in Refs. \cite{Martin:2001vx,Martin:2024qmi,McKeon:2004vs}.

The aim of this work is to find differential equations summing leading quantum corrections to an arbitrary Kähler potential. Also, it is necessary to solve these equations in those cases where it is possible to make a comparison with known examples (for example, the Wess-Zumino theory).

\section{Leading poles equation in the general chiral model}

The starting point of our analysis is the Bogoliubov-Parasiuk theorem \cite{BP,Hepp,Zimmermann,Breitenlohner:1975hg,Breitenlohner:1976te}. According to it, any local quantum theory has a remarkable property: in higher orders of perturbation theory after subtraction of divergent subgraphs in a divergent diagram, i.e. after performing an incomplete $\cal R$-operation (the $\cal R'$-operation) all remaining UV divergences must be local. Formulated in this form, the theorem allows us to relate the leading divergences in the first loop diagram with the $n$-loop diagrams \cite{Kazakov:2022pkc,Kazakov:2020kbj}. 

In our case for first-order computations, the relations between loops are not obvious, as one can encounter the following expressions:
\begin{equation}
\begin{gathered}
      k_1=-\frac{|w_2|^2}{2\gm^2},  \\
    2k_2= \frac{|w_2|^2}{\gm^3}\mathcal{D}_2k_1, \\ 3 k_3 = \frac{|w_2|^2}{\gm^3} \mathcal{D}_2 k_2-\frac{|w_2|^2}{2 \gm^4} \left(\mathcal{D}_2 k_1\right)^2,\\
    ... \text{higher loop orders}
\end{gathered}    
\end{equation}
where the covariant differential operator is $\mathcal{D}_2=\frac{\partial^2}{\partial \Phi \partial \bar{\Phi}}$. The resulting expressions bear little resemblance to the recurrence relations that typically arise in the study of non-renormalizable theories and do not help in constructing of differential equations.

It turns out that it is easier to find a differential equation for the effective Kähler potential using the so-called $\cal R$-rule \cite{Iakhibbaev:2026fst}. It states that differential equations for the sum of leading divergences can be written immediately, since the structure of differential equation for the sum of the leading poles is formally given as $$\mathbf{K}=\sum_{n=0}^\infty z^nk_n,$$ where $z=\lambda^2/\epsilon$ in the leading order has the structure of a single-loop diagram. In our case, we have to make a substitution which is justified due to the correspondence between one-loop diagram and its $\cal R$-operation \cite{Iakhibbaev:2026fst}
\beq
\frac{|w_2|^2}{2\gm^2}=-\frac{1}{2}\frac{|w_2|^2}{K_{\Phi \bar{\Phi}}^2} \rightarrow -\frac{1}{2}\frac{|w_2|^2}{(\mathcal{D}_2\bf{K})^2}.
\eeq
One can write the full differential equation as
\beq
\boxed{\frac{\partial}{\partial z} \textbf{K}=-\frac{1}{2}|w_2|^2 \left(\mathcal{D}_2\textbf{K}\right)^{-2}},\label{RGeq}
\eeq
with an initial condition $\textbf{K}(0;\bar{\Phi},\Phi)=K(\bar{\Phi},\Phi)$. This equation is a PDE, which seems to be unsolvable in general functions $K$ and $W$. 
It is possible to rewrite \eqref{RGeq} in a more suitable form acting by $\mathcal{D}_2$ on both sides
\begin{equation}
    \begin{gathered}
        \bm{G}^2\frac{\partial}{\partial z} \bm{G}=-\frac{1}{2}|w_3|^2 +\\+\bar{\bm{\Gamma}}\bar{w}_2 w_3+ \bm{\Gamma} w_2 \bar{w}_3+|w_2|^2 (\bm{R}-2|\bm{\Gamma}|^2),\label{RGeqG}
    \end{gathered}
\end{equation}
where $\bm{G}=\mathcal{D}_2\textbf{K}$ is the 'resummed' Kählerian metric, $\bm{\Gamma}$ is the corresponding connection and $\bm{R}$ is the Riemannian tensor on this manifold and the initial condition $\bm{G}(0,\Phi,\bar{\Phi})=\gm$. The convenience of this expression is that it is written exactly for the metric and allows one to track specific contributions from chiral superpotentials. It is interesting that in the case of a quadratic chiral superpotential $\sim m \Phi^2+h \Phi + \Lambda$ (i.e. for $w_3=\bar{w}_3=0$), the latter equation can be simplified to
\begin{equation}
        \bm{G}^2\frac{\partial}{\partial z} \bm{G}=|m|^2 (\bm{R}-2|\bm{\Gamma}|^2),\label{RGeqG2}
\end{equation}
which has some similarities with the Kähler-Ricci flow equation. 

In order to obtain the effective Kähler potential from \eqref{RGeq}, it is necessary to replace the singularities with logarithms as 
\begin{equation}
    \textbf{K}(z, \Phi,\bar{\Phi}) \rightarrow \textbf{K}\left(\lambda \log\left(\frac{\gm^2 \mu^2}{|w_2|^2}\right),\Phi,\bar{\Phi}\right),
\end{equation}
i.e., the substitution follows the same logic as in Ref. \cite{Kazakov:2022pkc}.

Equations \eqref{RGeq} and \eqref{RGeqG} are generalizations of the usual renormalization group equation in the case of non-renormalizable theory and contains all the information about the behavior of leading divergences and logarithms. In general, partial derivative equations of this type cannot be solved \cite{Kazakov:2022pkc}. In the next section, we apply this equation to some simple models and show its validity. To proceed, we solve the equation analytically or numerically.

\section{Analysis of some models}
\subsection{Wess-Zumino model}

In the case of the Wess-Zumino model, the ansatz should be of the following form
\beq
\textbf{K}_{WZ}(z;\bar{\Phi},\Phi)=\bar{\Phi}\Phi ~\kappa_{WZ}(z),
\eeq
this substitution is required by the dimensional considerations. In this case $\gm=1$ and Eq. \eqref{RGeq} is considerably simplified. So the ansatz turns \eqref{RGeq} to the well-known ordinary differential equation
\beq
\kappa_{WZ} '(z)=-\frac{1}{2}\kappa_{WZ} (z)^{-2}, \label{RGeqWZ}
\eeq
with the initial condition $\kappa_{WZ}(0)=1$. This equation can be integrated to give 
\beq
\kappa_{WZ}(z)=\left(1-\frac{3}{2}z\right)^{1/3}, \label{eq::WZ3sol}
\eeq
which gives exactly all the leading divergences in the Wess-Zumino model and reproduces the relevant coefficient of the one-loop $\beta$-function (anomalous dimension in the one-loop case is $\gamma_1=1/2$) and the running coupling in the latter theory \cite{Huq:1977eu,Fogleman:1983hm,Martin:2024qmi}:
\beq
\lambda^2_{WZ}(\mu)=\frac{\lambda^2}{1+\frac{3}{2}\frac{\lambda^2}{16\pi^2}\log(\frac{\bar{\Phi}\Phi}{\mu^2})}.
\eeq
The validity of PT requires that the expansion parameter obeys $\lambda^2/16\pi^2<1$, and the validity of the leading logarithmic approximation dictates $\log(\frac{\bar{\Phi}\Phi}{\mu^2})>1$.  The occurrence of the Landau pole should be mentioned: it indicates a violation of the convergence conditions of the series. 

This example proves the correctness of the obtained equation. To proceed, one has to consider more general models.

\subsection{Wess-Zumino model with an arbitrary power-like interaction}

From dimensional considerations, the following ansatz arises for the non-renormalizable theory with the chiral superpotential $W=\Phi^p/p!$
\beq
\textbf{K}(z;\bar{\Phi},\Phi)=\bar{\Phi}\Phi ~\kappa(z (\bar{\Phi}\Phi)^{p-3}),
\eeq
so for the sake of convenience, we introduce the variables $\tau=z (\bar{\Phi}\Phi)^{p-3}$ and $\xi=p-3$; therefore one can simplify the general equation \ref{RGeq} to the following nonlinear ODE:
\beq
\begin{gathered}
\kappa'(\tau)=\frac{-\Gamma (2+\xi )^{-2}}{2 \left(\kappa (\tau )+\xi  \left((\xi -2) \tau\kappa '(\tau )+\xi  \tau^2  \kappa ''(\tau )\right)\right)^2},
\end{gathered}\label{eq::WZgeneq}
\eeq
with the initial condition $\kappa(0)=1$ and $\kappa'(0)=-\frac{1}{2 \Gamma (\xi+2)^2}$. Note that for $\xi=0$ (or $p=3$) we have the solution given in \eqref{eq::WZ3sol}. 

One can find that \eqref{eq::WZgeneq} is invariant under the following scale symmetry:
\begin{equation}
    \tau \rightarrow a^p \tau, ~\kappa \rightarrow a^{p/3} \kappa,
\end{equation}
where $a$ is a constant. Using this transformation, we can construct an invariant structure which reads off
\begin{equation}
    u = \tau^{-1/3} \kappa(\tau) ,~v=\tau^{2/3} \kappa'(\tau).
\end{equation}
It should be noted here that the scale symmetry of the equation is broken by the initial conditions, which choose the sign for the derivative of the solution at $\tau=0$, i.e. singular or discontinuous behavior is expected.

After  substituting \eqref{eq::WZgeneq}, the order is reduced, and the result is a first order ODE
\begin{equation}
\begin{gathered}
    \left(\xi^2 v v'+\frac{2}{3}\xi(\xi-3)v+(\xi-3)^2 \frac{u}{9}\right)^2= -\frac{\hat{\alpha}^2}{v-u/3},\label{nOde}
\end{gathered}
\end{equation}
where $\hat \alpha^2=\frac{1}{2}\Gamma(\xi+2)^{-2}$. One can re-express \eqref{eq::WZgeneq} as an autonomous ODE and treat the system as a nonlinear dynamical system.  The Jacobian matrix for the ODE \eqref{nOde} can be found in the following form $\text{tr}(J)=(3-\xi) (5\xi-3)/6\xi^2$. It vanishes at $\xi=3/5$ and $\xi=3$. The first point is the supercritical Andronov-Hopf bifurcation point, and at $3/5 < \xi <3$,  the limit cycle appears so that solutions have log-periodic behavior. It is interesting that at the point $\xi=3$ the equation is integrable and the parametric solution can be represented as a combination of Airy functions and their derivatives \cite{zaitsev2002handbook}.  This solution scales as $\kappa_{\xi=3}(\tau)\sim a_1 \tau^{1/3}+a_2 \tau^{-1/3}$ where $a_i$ are some constants defined by the initial conditions. These solutions can be analyzed in detail with the help of built-in packages in \texttt{Maple}.  

To proceed to study the small $\xi$ behavior, we linearize \eqref{eq::WZgeneq} and analyze its near-critical behavior. The procedure results the following equation:
\beq
\frac{1}{\kappa (\tau )^2}-\frac{4 \xi  \left(\tau  \kappa '(\tau )\right)}{\kappa (\tau )^3}=\kappa '(\tau )\hat \alpha^2, \label{linRGeq}
\eeq
 It can be seen that in the limit $\xi \rightarrow 0 $ we recover Eq \eqref{RGeqWZ}.
In the small $\xi$ limit, the equation can be represented as a simple algebraic equation
\beq
\tau =\hat\alpha^2\frac{\kappa (\tau )^3-\kappa (\tau )^{4 \xi }}{ (4 \xi -3)}\hat .
\eeq
In some particular cases, this equation can be solved. For example, one can choose $\xi=1/4$, which is in the applicability limit of the approximation and gets immediately conjugate solutions
\beq
\kappa (\bar\tau )= \frac{\left(\sqrt{ \bar\tau^2-1/3}- \bar\tau \right)^{2/3}+1}{3^{1/3} (\sqrt{ \bar\tau-1/3}- \bar\tau )^{1/3}},
\eeq
where $\bar\tau=-\frac{\tau}{2} \Gamma\left(\frac{9}{4}\right)^{-2}$. A similar solution can be found for $\xi=1/2$
\begin{equation}
    \kappa(\hat\tau)=\frac{1}{3} \left({f(\hat\tau)}^{1/3}+{{f(\hat\tau)}}^{-1/3}+1\right),
\end{equation}
where $f(\tau)=1+\frac{27 \hat\tau}{2}+3 \sqrt{3} \sqrt{\hat\tau  \left(\frac{27 \hat\tau }{4}+1\right)}$, where $\hat\tau=-\frac{\tau}{2}\Gamma(5/2)^{-2}$.  Note that the special point $\xi=3/4$ is a signal of violation of the approximation.

\begin{figure}
    \centering
    \includegraphics[width=0.5\linewidth]{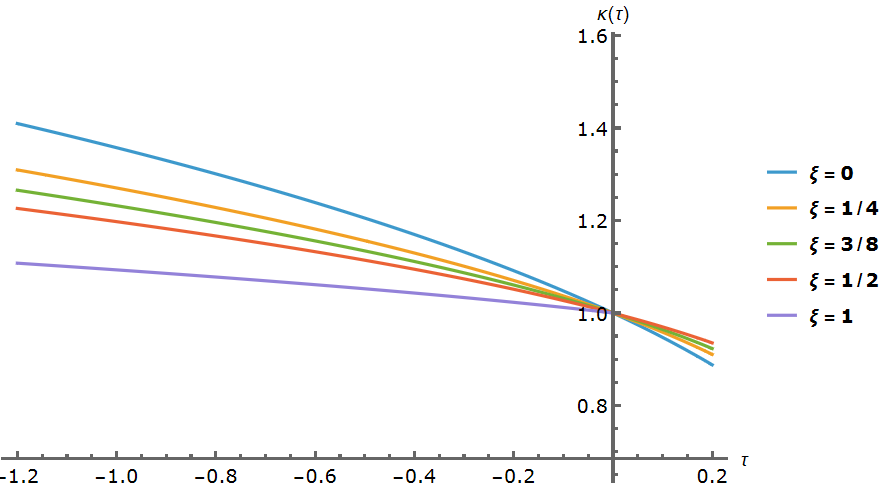}
    \caption{Solution of equations $\kappa(\tau)$ for some parameters $\xi$. For the exact solution at $\xi=1$ a finite discountinuity of the solution appears near $\tau=0$.}
    \label{fig:kappaplot}
\end{figure}

\begin{figure}
    \centering
    \includegraphics[width=0.5\linewidth]{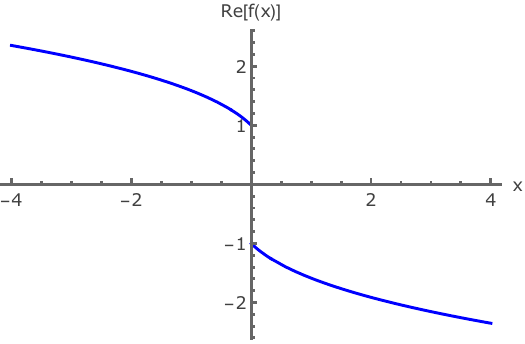}
    \caption{Typical solutions $f(x)$ of ODEs \eqref{eq::WZgeneq} and \eqref{flatLogeq}. Appearance of the discontinuity at $x=0$ is depicted.}
    \label{fig:typbehaviour}
\end{figure}

Outside small $\xi$ approximation, we notice a discontinuity of the function at zero (see Fig. \ref{fig:typbehaviour}), which is similar to that found in the effective potentials of general scalar models \cite{Kazakov:2022pkc}. One can see that in general all solutions are very close, they scale universally at large $\tau$ as $\sim \kappa_0\tau ^{1/3}$ and very similar to the standard RG solution of the renormalizable Wess-Zumino model.

\subsection{Simple 'no-scale' chiral model}

Let us consider the classical no-scale Kähler potential $K_0(\Phi,\bar{\Phi})=-3 \Lambda^2 \log\left(\frac{\Phi+\bar{\Phi}}{\Lambda}\right)$ which corresponds to constant negative curvature geometry $R=-\frac{2}{3\Lambda^2}$. Models like these originate from superstring dimensional reduction \cite{Witten:1985xb}. They are frequently used in inflationary cosmology in various forms since they provide many convenient properties: flatness of inflationary potential, a natural solution to the $\eta$-problem etc. Moreover the no-scale models can
yield an inflation potential with a necessary slow-roll plateau, and they are often considered as successful in cosmological phenomenology   \cite{Nanopoulos:1982bv,Ketov:2014qha,Moursy:2020sit,Ellis:2013xoa,Ellis:2026ceb}.

The Kählerian metric of this model can be represented as $\gm=3e^{2K_0/3\Lambda^2}$ and the connection is $\gamma=-\frac{1}{\Lambda} e^{K_0/3\Lambda^2}$. For  simplicity, we introduce the massive chiral superpotential $W(\Phi)=\Lambda \Phi^2/2$. 
Note that in the loop expansion of this model there are no sunset-type diagrams generated by $w_3$, all diagrams are proportional to $|w_2|^2=\Lambda^2$ (see also \eqref{RGeqG2}). Inserting these data into \eqref{oneloop} and \eqref{twoloop},  we obtain that the effective Kähler potential has the following form
\begin{equation}
   \mathbf{K}=K_0+\frac{\Lambda^{2}}{9}\sum_{j=1}^\infty t^j~\mathbf{K}_j=K_0+\frac{\Lambda^{2}}{9} \varrho(t), \label{subscale}
\end{equation}
where $t= \mathfrak{g}^2 /{9 \epsilon }$ (assuming $g^2/9<1$ and $\log\left(\frac{\gm^2\mu^2}{ \Lambda^2}\right)>1$ for the convergence of PT series).  After substitution of \eqref{subscale} into \eqref{RGeq} or \eqref{RGeqG}
one can simplify the equation as follows:
\begin{equation}
 \varrho'=-\frac{729}{2(27-12t \varrho'+16 t^2 \varrho'')^2} \label{flatLogeq},
\end{equation}
with the initial conditions $\varrho(0)=0,\varrho'(0)=-1/2$.
This equation is similar to the previous ones and has the same scale symmetry. Also, Eq. \eqref{flatLogeq} admits reduction of the order and allows for a numerical or qualitative study (see Fig. \ref{fig:typbehaviour}). Note also that the solutions are always covariant. This equation has the same symmetries and has the same scaling at large $t$ as in $\varrho \sim \varrho_0 ~t^{1/3}$. 

\section{Conclusions}

In this work, we have calculated loop corrections for up to three loops, which is consistent with the result known in the literature from Ref. \cite{GrootNibbelink:2005nez}. Based on the correspondence between loops and $\cal R$-operation, we constructed the equation describing the leading singular and logarithmic corrections to the effective Kähler potential for an arbitrary theory with any classical Kähler and chiral superpotentials. In general, this is a partial differential equation, the solution of which is most likely impossible. Nevertheless, some models allow us to reduce the problem to studying of ordinary differential equations. In this paper, we demonstrated that the resulting equation reproduces the case of the renormalizable Wess-Zumino model. We also managed to study some examples of non-renormalizable potentials near the critical dimension $p=3+\xi$ and investigated the behavior of the simplest model with the constant curvature logarithmic Kähler potential. The universal power-like behavior is established for both examples. 

The obtained equations \eqref{RGeq} and  \eqref{RGeqG} have a simple structure; it is important to find examples of some models for which the original PDEs are integrable.
It would also be interesting to derive and study the equations in the next-to-leading logarithmic approximation and track the dependence of solutions on the subtraction scheme as well as to find the corresponding equations for the general chiral model \eqref{GenLagr} in a curved supergravity background.

It is known that in the RG equations, there is a simple recurrence relation between the loop orders in the Wess-Zumino theory due to the connection between the beta function and the anomalous dimension \cite{Sen:1981hk,Avdeev:1982jx,Jack:1996qq}. It would be interesting to discover a similar connection in our generalized approach or a connection with Ref. \cite{Lakhal:2025nbh}. It would additionally be interesting to obtain analogous equations for the case of renormalisable and non-renormalisable supersymmetric gauge theories \textcolor{black}{and especially extended supersymmetric models}. 

The results of this work can be applied in a wide range of practical problems, particularly in string-inspired theories of cosmological inflation, where the study of loop corrections to Kähler potentials is often important (see, for instance, work on inflationary cosmological models \cite{Nanopoulos:1982bv,Ellis:2013xoa} or more relevant study where loop corrections are considered \cite{Ellis:2026ceb}).

\subsection*{Acknowledgements}

R.M. Iakhibbaev and  D.M. Tolkachev thank the CERN Theoretical Physics Department for their hospitality during the preparation of this work. The authors are grateful to  D.V. Bykov, D.I. Kazakov, S.V. Ketov and for useful comments and discussions. The authors are thankful to I.L. Buchbinder for numerous discussions and attracting attention to Refs. \cite{Buchbinder:1999sk,Buchbinder:1999jw,Buchbinder:1999ui}.   A. I. Mukhaeva’s work is supported by the Foundation for the Advancement of Theoretical Physics and Mathematics BASIS, No 24-1-4-36-1.

\bibliography{refs}
\bibliographystyle{unsrt}

\end{document}